\documentclass[a4paper,10pt]{article}
\usepackage[utf8]{inputenc}
\usepackage{geometry}
\usepackage{natbib}
\usepackage[svgnames]{xcolor} 

\usepackage{multicol} 
\usepackage[svgnames]{xcolor} 

\usepackage{times} 

\usepackage{graphicx} 
\usepackage{graphicx}
\usepackage{gensymb}
\usepackage{lscape}
\graphicspath{{figures/}} 
\usepackage{booktabs} 
\usepackage[font=small,labelfont=bf]{caption} 
\usepackage{amsfonts, amsmath, amsthm, amssymb} 
\usepackage{wrapfig} 
\usepackage{graphicx} 
\usepackage{gensymb}
\usepackage{colortbl}
\usepackage{lscape}
\graphicspath{{figures/}} 
\usepackage[font=small,labelfont=bf]{caption} 
\usepackage{amsfonts, amsmath, amsthm, amssymb} 
\usepackage{wrapfig} 
\usepackage{authblk}
\usepackage{setspace}
\usepackage{lineno}

\title{LaRa after RISE: expected improvement in the Mars rotation and interior models.}
\author[1]{M.-J. P\'eters}
\author[1]{S. Le Maistre} 
\author[1]{M. Yseboodt}
\author[2]{J.-C. Marty}
\author[1]{A. Rivoldini}
\author[1]{T. Van Hoolst}
\author[1]{V. Dehant}
\affil[1]{\small Royal Observatory of Belgium, Brussels (Belgium)}
\affil[2]{\small CNES, Toulouse (France)}
\date{}

\begin{document}

\maketitle

\begin{center}
 Accepted in PSS, October 2019
\end{center}
\vspace{0.5cm}

\begin{abstract}
Two years after InSight's arrival at Mars, the ExoMars 2020 mission will land on the opposite side of the red planet. Similarly to InSight, which carries the RISE (Rotation and Interior Structure Experiment) radio-science experiment, the ExoMars mission will have on board the Lander Radio-science (LaRa) experiment. The X-band transponders on RISE and LaRa, allowing for direct radio-link between the landers and stations on Earth, are dedicated to the investigation of Mars' deep interior through the precise measurement of the planet's rotation and orientation. The benefit of having LaRa after RISE for the determination of the Mars orientation and rotation parameters is demonstrated and the resulting improved constraints on the interior structure of Mars and, in particular, on its core are quantified via numerical simulations. In particular, we show that the amplitudes of the semi-annual prograde ($p_2)$ and the ter-annual retrograde ($r_3$) nutations will be determined with a precision of 6 and 4 milliarcseconds respectively by combining 700 days of RISE data with 700 days of LaRa data, about 35$\%$ more precise than what is expected from RISE alone. The impact of such an improvement on the determination of the core size of Mars is discussed and shown to be significant. 

\end{abstract}

\section{Introduction}

The NASA InSight stationary lander was launched in May 2018 and landed on Mars on November 26, 2018. InSight's payload consists of three main geophysical instruments including the radio-science experiment RISE (\cite{Folkner:2018}). ExoMars 2020 is an ESA-ROSCOSMOS joint mission that consists of a European rover and a Russian platform that will be sent to the surface of Mars. The latter will carry 13 instruments including the LaRa radio-science experiment (\cite{dehant:2019}). The launch is scheduled for July 2020 and the arrival at Mars is expected in March 2021. \\
\\
The LaRa and RISE instruments are both X-band (8.4 GHz) coherent transponders designed to determine Mars' Orientation and rotation Parameters (MOP) by analyzing the Doppler shift of the two-way radio signals between the lander on Mars and ground stations on Earth (e.g. \citet{folkner:1997}). The targeted MOP are the variations of Mars' rotation rate or length-of-day (LOD) and the variations of the orientation of the rotation axis in space (precession and nutations). In order to infer knowledge about the interior structure and in particular about the core, the landers will be tracked with high accuracy. In the past, similar radio-science experiments have been carried out during the Viking (\cite{yoder:1997}) and Mars Pathfinder missions (\cite{folkner:1997}) to study the rotation of Mars but they were not sufficiently accurate to deduce any meaningful constraints on the core parameters. Rovers and orbiters have later been used to determine the MOP from Doppler tracking (\cite{kuchynka:2014}, \cite{konopliv:2016}, \cite{lemaistre:2013}). Only \cite{lemaistre:2013} and \cite{lemaistre:2019} estimated the nutation amplitudes although the uncertainties remained too large to significantly constrain the interior structure of Mars.\\
\\
In this paper we investigate the improvement on the determination of the Mars orientation and rotation parameters that can be achieved through the synergy between LaRa and RISE, onboard the ExoMars 2020 \citep{dehant:2019} and the ongoing InSight missions \citep{Folkner:2018}. We perform numerical simulations to quantify the improvement on the MOP that will be achieved by combining RISE and LaRa data. This combined solution is compared to the solutions obtained by the individual experiments. For that, we use the GINS (G\'eod\'esie par Int\'egrations Num\'eriques Simultan\'ees) software developed by Centre d'\'Etudes Spatiales (CNES) and further adapted at Royal Observatory of Belgium for planetary geodesy applications.\\
\\
In Section~\ref{sec:setup}, we present the mission scenarios considered to perform our simulations which depend on the actual mission's operational parameters. In Section~\ref{sec:rotmod}, we introduce the Martian rotation model used to simulate the Doppler data. In Section~\ref{sec:res}, the results of our simulations are showed and we discuss the expected constraints on the core in Section~\ref{sec:interior}. Conclusions are presented in~\ref{sec:conclusion}.

\section{Simulations setup}
\label{sec:setup}
The operational characteristics considered for each mission and used to create the simulated data are given in Table~\ref{table:scenario}. For both missions, the tracking is assumed to be performed from ground stations at Goldstone (USA), Madrid (Spain) and Canberra (Australia). These stations belong to the NASA Deep Space Network (DSN). Their locations in the inertial space are well known. On Mars, InSight-RISE is located at latitude 4.5$^\circ$N, longitude 135.62$^\circ$E, altitude -2.6 km in Elysium Planitia, coordinates which were estimated right after landing by \cite{Parker:2019aa}. LaRa will be positioned at Oxia Planum (latitude 18.3$\degree$N, longitude 335.37$\degree$E, altitude -2 km), the ExoMars 2020 selected landing site on the other side of the planet opposite Elysium Planitia. For both experiments we consider 700 days of operations (more or less the nominal operation time), starting Nov. 27th, 2018 for RISE and March 19th, 2021 for LaRa. We generate synthetic data sets consisting of two-way X-band Doppler measurements that we have affected by a white noise of 0.05 mm/s at 60s integration time, which is typical for the Martian landed missions (e.g. \cite{dehant:2019}). Although \cite{Folkner:2018} used a more realistic noise model, which takes into account the solar plasma and Earth troposphere, we show below that our simplified noise model predicts MOP uncertainties consistent with those obtained by \cite{Folkner:2018} with the more realistic noise. Moreover, we removed data points acquired at Sun-Earth-Probe (SEP) angles smaller than 15$\degree$ as will be done in the true data analysis since at those angles interplanetary plasma strongly perturbs the radio signal. These blackout periods around solar conjunctions have a small impact on the estimated uncertainties of nutation amplitudes (\cite{lemaistre:2019}).\\
\\
The RISE and LaRa experiments have different antenna configurations. LaRa has one receiving (Rx) and two (for redundancy reason) transmitting (Tx) monopole antennas. They provide a maximal gain between 30$\degree$ and 55$^\circ$ of Line-Of-Sight (LOS) elevation in every azimuth (\cite{Karki:2019aa}). LaRa's simulated Doppler data last 45 minutes\footnote{We reduced the hour-long nominal tracking pass by 15 minutes to account for possible long acquisition time (worst case). Note that tracking LaRa for more than one hour is compromised due to the limited power available one the lander.} per session, with two sessions per week. Tracking is performed when Earth LOS is within a range of [35$^\circ$, 45$^\circ$] of elevation above the local horizon of the lander. LaRa is tracked with a LOS to the Earth switching every session from East to West azimuth direction and reciprocally (\cite{dehant:2019}).
RISE has two medium-gain horn antennas able to receive and transmit the signal at the same time but only with one antenna at a time since one points eastward and the other one points westward. The antenna's maximum gain is at 30$\degree$ elevation above the lander, with an aperture of about 25$\degree$ off bore sight. RISE is nominally tracked each day for about an hour when the Earth is above 10$\degree$ of elevation (to avoid potential obstacles and ground effects on the signal at lower elevation), but still in the lower part of the antenna lobe, i.e. $\le 30\degree$ elevation, in order to get the best accuracy for estimation of the moment of inertia of the core (\cite{Folkner:2018}). Although nominal operation foresees alternate tracking between the antennas, InSight has only been tracked during the morning hours (Eastward antenna) of the first $\sim$200 days because of operational constraints (i.e. daily command uploading). This is taken into account in our simulations, which are divided in two scenarios of RISE operations: the first one, called 'optimal' hereafter, considers antennas pointing to 15.5$^\circ$ south of due east and to 6$^\circ$ north of due west as planned before landing (\cite{Folkner:2018}). In this scenario, tracking windows are alternately centred on these directions (except for the first 200 days). The second scenario, called 'nominal', takes into account the 5-degree-clockwise\footnote{When viewed from above} azimuth offset of RISE antennas (Folkner pers. com.), and enables RISE tracking only for azimuth within $\pm 25\degree$ from the actual antennas bore sight, preventing tracking outside the antennas aperture, and is therefore close to reality. These two specific aspects of the 'nominal' case preclude systematic east-west alternative tracking and reduce the total number of days of operation with respect to the 'optimal' case, degrading the experiment performances as shown below and further explained in \cite{lemaistre:2019}.

\renewcommand{\arraystretch}{1.2}
\setlength{\tabcolsep}{0.5cm}
\begin{center}
\begin{tabular}{c|cc}
& \large{\bf{RISE}}  & {\large{\bf{LaRa}}} \\
 \hline \hline
 \multicolumn{1}{c|}{Mission}              & InSight                       &  ExoMars 2020 \\
 \hline
\multicolumn{1}{c|}{Landing site}           & Elysium Planitia   &  Oxia Planum \\
  & (4.5$^\circ$N, 135.62$^\circ$E, -2.6 km)  &   (18.3$\degree$N, 335.37$\degree$E, -2 km)\\
\hline
\multicolumn{1}{c|}{Starting date}              & November 2018                        &  Mars 2021 \\
\hline
\multicolumn{1}{c|}{Mission duration}           & 700 days  &  700 days \\
\hline
\multicolumn{1}{c|}{Tracking pass duration}   & 60 min/day                             & 45 min twice per week \\
\hline
\multicolumn{1}{c|}{Total number of data points}   & 36 795                                & 11 805 \\
\hline
\multicolumn{1}{c|}{LOS elevation}  & [10$\degree$, 30$\degree$]          & [35$\degree$, 45$\degree$] \\ 
\hline
\multicolumn{1}{c|}{Doppler noise}              & \multicolumn{2}{c} {White noise of 0.05 mm/s for 60~s of integration time} \\
\hline
\multicolumn{1}{c|}{Ground stations}            & \multicolumn{2}{c} {Deep Space Network stations (DSS-15,45,65)} \\
\end{tabular}%
\end{center}
\captionof{table}{RISE and LaRa operation characteristics considered in our simulations.}
\label{table:scenario}
\vspace{0.3cm}

\begin{figure}[h!] 
\begin{center} 
\includegraphics[scale=0.5]{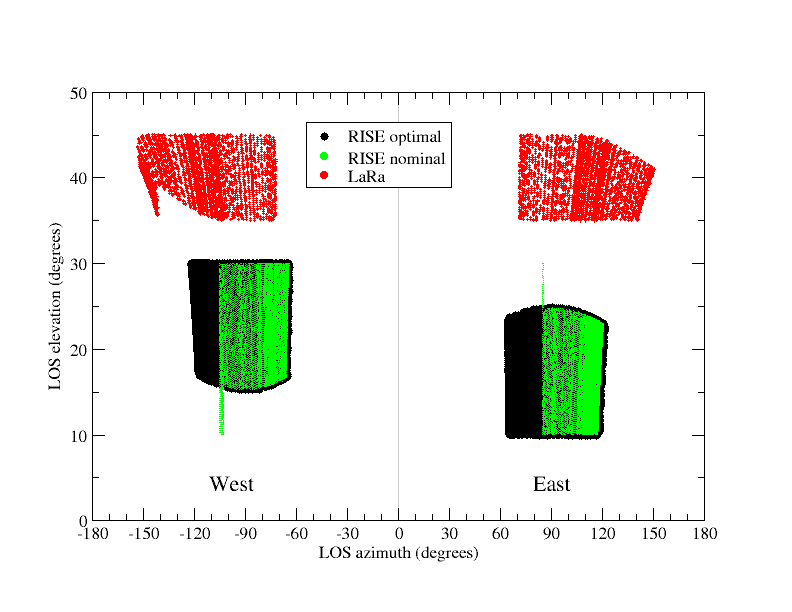}
\caption {Earth LOS direction in terms of azimuth and elevation for both transponders. LaRa is in red and the different RISE configurations are the optimal case in black and the nominal case in green.} \label{fig:azi_elev}
\end{center}
\end{figure}

\begin{figure}[h!]
\begin{center} 
\includegraphics[scale=0.5]{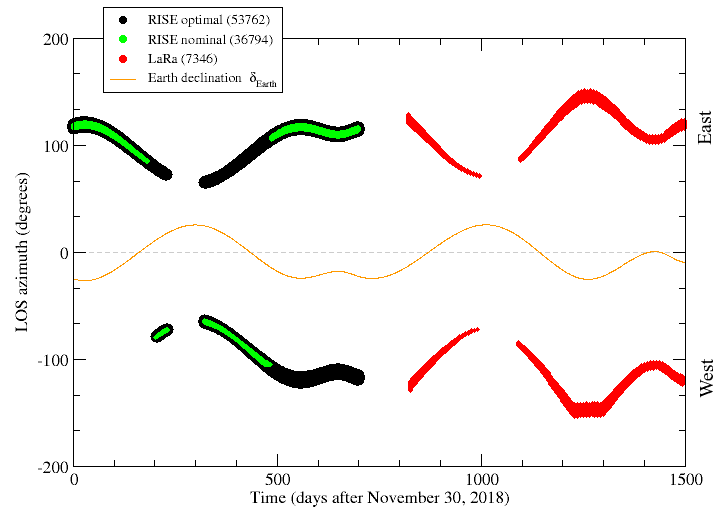}
\caption {Temporal evolution of Earth LOS azimuth for both transponders. LaRa is in red, RISE optimal case is in black and RISE nominal case is in green. The orange line shows the Earth declination $\delta_{Earth}$ (in degrees). The number of data points is indicated in brackets.} \label{fig:azi}
\end{center}
\end{figure}

\begin{figure}[h!]
\begin{center} 
\includegraphics[scale=0.5]{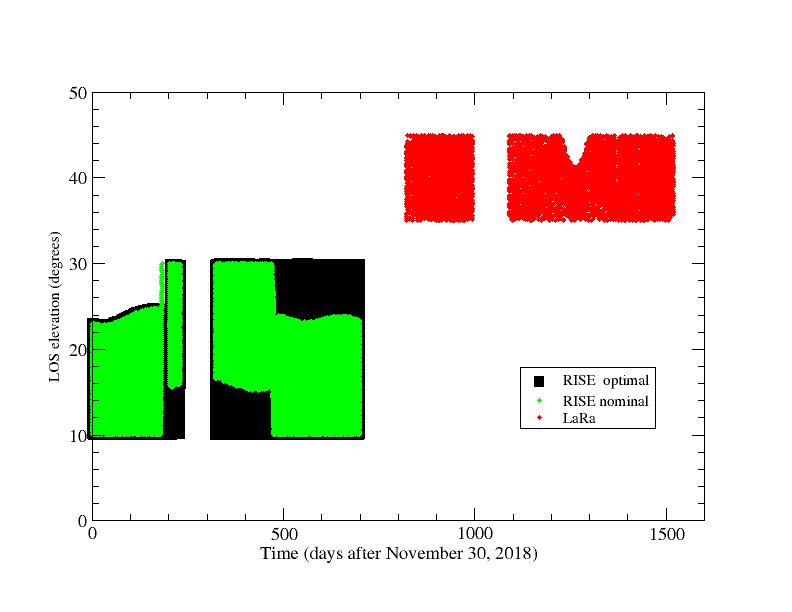}
\caption {Temporal evolution of Earth LOS elevation for both transponders. LaRa is in red, RISE optimal case is in black and RISE nominal case is in green.} \label{fig:elev}
\end{center}
\end{figure}

\begin{figure}[h!]
\begin{center} 
\includegraphics[scale=0.5]{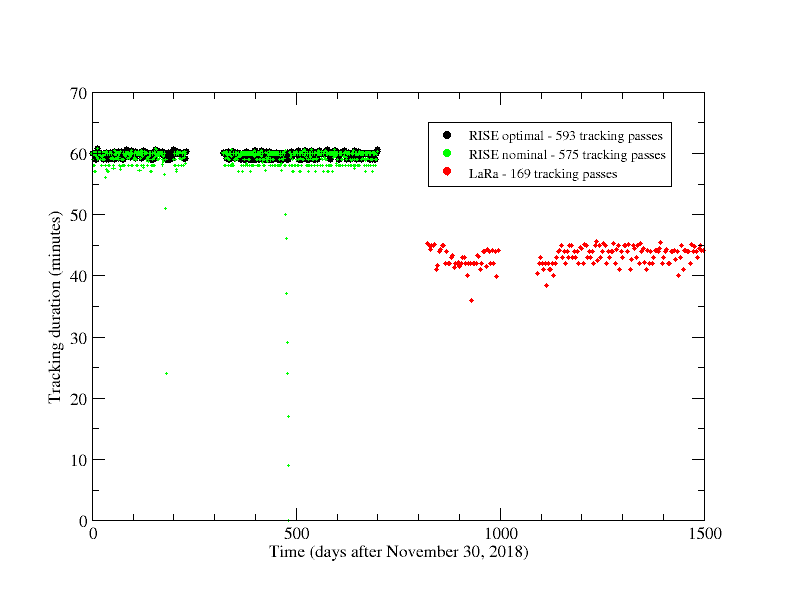}
\caption {Temporal evolution of the tracking duration for LaRa, RISE optimal and RISE nominal .} \label{fig:track}
\end{center}
\end{figure}

We show the Earth LOS geometry during tracking in Figures~\ref{fig:azi_elev} to \ref{fig:elev} and show the tracking session duration for both LaRa and RISE as considered in our simulations in Fig.~\ref{fig:track}. These quantities crucially because they affect the sensitivity of the Doppler measurements to the MOP \citep{yseboodt:2017,lemaistre:2019}. \\
\\
Fig.~\ref{fig:azi_elev} represents the Earth LOS direction in terms of elevation and azimuth in the sky of both landers. When considering the two successive missions, the figure shows a pretty diversified geometry, with azimuth ranging in $[60^\circ;150^\circ]$ and $[-150^\circ;-60^\circ]$, and elevations ranging from 10$^\circ$ to 45$^\circ$. Separately, LaRa azimuths are within [-153$^\circ$,-72$^\circ$] and [71$^\circ$,151$^\circ$] and RISE between [-125$^\circ$,-65$^\circ$] and [62$^\circ$, 123$^\circ$]. LaRa covers a larger range of azimuth. Moreover, the RISE nominal scenario (green) has a strongly reduced coverage in azimuth compared to the optimal scenario (black), which mostly results from the slight off-pointing of 5 degrees in the nominal case. \\
\\
LaRa tracking systematically switches from East to West and West to East over the whole mission (Fig.~\ref{fig:azi}). This alternation is also true for the RISE optimal scenario, but only after 200 days of observation, while the 'nominal' configuration does mostly not allow this reciprocation, making the Earth visible by only one of RISE antennas for most of the time. The time evolution of the Earth declination $\delta_{Earth}$ (angle between the Mars-Earth vector and the Martian equatorial plane) is also shown in Fig.~\ref{fig:azi}. When $\delta_{Earth}$ is close to zero, the sensitivity of the DTE (direct-to-Earth) Doppler  measurements to the nutation is null because the rotation axis of Mars is then perpendicular to the LOS \citep{yseboodt:2017}. Unfortunately, the Earth declination maxima, when DTE Doppler sensitivity to nutation is maximum, are close to the solar conjunctions when data are discarded because of large plasma noise (i.e. when $SEP <15\degree$). \\
\\
Fig.~\ref{fig:elev} represents the LOS elevation as a function of time. For the RISE nominal case, the range of elevation is limited with respect to the optimal one to $[15^\circ;30^\circ]$ between 200 and 450 days of observation and to $[10^\circ;25^\circ]$ afterwards due to the azimuthal off-pointing of 5 degrees. During the first 100 days of RISE tracking, the maximum LOS elevation does not reach 30 degrees due to our simulation data filtering. The most important is that the global elevation range should be conserved. During the LaRa tracking, the Earth is between 35 and 45 degrees above the horizon. The maximum LOS elevation drops to about 40 degrees for hundred days because of the low Earth elevation in the lander sky during the Martian winter at Oxia Planum. \\
\\
The tracking duration remains very close to 60 minutes at all times for the RISE optimal case and can drop by a few minutes for the nominal case mainly due to azimuth cutoff (see Fig.~\ref{fig:track}). Some variations of about 5 minutes occur for LaRa by construction of our simulated data set. With a 60-min track per day, RISE accumulates 5 to 7 times more data than LaRa over one Martian year, but the diversity in LaRa Doppler geometry appears to be more favorable for the determination of some parameters like the nutation parameters. We will show the impact of these antagonist effects on the estimation of the nutation parameters in the section 4 (see further details in \cite{lemaistre:2019}).

\vspace{1cm}

\clearpage
\section{Martian Rotation model}
\label{sec:rotmod}
In order to build synthetic Doppler data we use the rotation model described in \cite{lemaistre:2012,lemaistre:2019}. The rotation model is based on the pro-- and retrograde nutation amplitudes of Mars assumed to behave as a rigid body \citep{Roosbeek:1999} (rescaled with the latest value of the precession constant). The latter are augmented with a model for the non-rigid planet with a fluid core. For the precession and spin angle variations, we follow \cite{konopliv:2016} and for the polar motion we use the model of \cite{vandenacker:2002}. The parameters governing the rotation model are given in Table \ref{table:rotation_model}. 

In our simulations, we focus on the nutation parameters because they provide important constraints on the interior of Mars (e.g. \cite{vanhoolst:2015}). These parameters are the amplitudes of the prograde $p_m$ and retrograde $r_m$ nutation terms respectively and are called hereafter the ``nutation parameters'' (\cite{defraigne:1995}).
$m=1$ corresponds to an annual Mars orbital period of 686.96 Earth days and the following $m=2,3,4...$ correspond to the fractions: semi-annual (343.5 days), ter-annual (229 days), the 4-annual (172 days), etc. \\
\\
The nutation amplitudes depend on the interior structure and are particularly sensitive to the core and its physical state (liquid or solid). They can be significantly amplified when in near resonance with the Free Core Nutation (FCN), one of the rotational normal modes of Mars (\cite{dehant:2015aa}). The non-rigid nutations amplitudes can be expressed according to (\cite{sasao:1980}):
\begin{equation}
p'_m=  p_m \left( 1+F\frac{\sigma_m}{\sigma_m-\sigma_0 } \right)  \qquad r'_m=  r_m \left( 1+F\frac{\sigma_m}{\sigma_m+\sigma_0 } \right)\end{equation}
where the terms between parentheses are the transfer functions that relate the nutation amplitudes of a rigid planet ($p_m$, $r_m$) to those of a non-rigid one ($p'_m$, $r'_m$). 
The core momentum F and the FCN frequency $\sigma_0$ in (1) are expressed as \begin{equation}
\sigma_0=-\Omega\frac{A}{A-A_f}(e_f-\beta)\quad \textrm{and}\quad F=\frac{C_f}{C-C_f}(1-\frac{\gamma}{e})
\end{equation}

with $A, C$ are equatorial and polar principal moments of inertia, of Mars $A_f, C_f$ are the principal moments of inertia of the fluid core, $e$ is the dynamical flattening ($e=C-A/A$), $e_f$ is the dynamical flattening of the fluid core and $\Omega$ is the mean rotation rate. $\gamma$ and $\beta$ are compliances that characterize the core's capacity to deform through rotation rate variation and tidal forcing \citep{sasao:1980}. \\
\\
Following the method of \citet{lemaistre:2012,lemaistre:2019}, we estimate the lander position coordinates (X,Y,Z), the precession rate  $\dot{\psi_0}$, the obliquity rate $\dot{I_0}$, the annual and semi-annual LOD variations amplitudes and the annual, semi-annual and ter-annual nutation amplitudes. Other MOP reported in Table \ref{table:rotation_model} are fixed because they have a very small signature in the Doppler observable (\cite{yseboodt:2017}) or because they are considered as well-known\footnote{The current level of uncertainty on these parameters propagates to the Doppler signal at a level significantly lower than the noise floor.}. Synthetic Doppler data are created using a model of polar motion (\cite{vandenacker:2002}) but those parameters are not estimated because the DTE Doppler data from equatorial landers are not sensitive to the polar motion (\cite{yseboodt:2017}). However, the Chandler Wobble amplitude might be large enough to be detected by LaRa (\cite{dehant:2019}) but this is neglected here since it does not inform on the interior.\\
\\

\renewcommand{\arraystretch}{1.2}

\begin{table}[h!]
\begin{center}
\begin{tabular}{c c c c c}
\hline
\hline
   \textbf{Symbol (unit)} & \textbf{Parameter} & \textbf{Nominal value} & \textbf{A priori constraint ($1\sigma$)} \\
  \hline 
  \hline
 $\psi_0$ (deg)  & Longitude of the node at J2000 & 81.97 & fixed \\
 $I_0$ (deg) & Obliquity at J2000 & 25.189 & fixed \\
 $\phi_0$ (deg) & Position of Prime meridian at J2000 & 133.386 & fixed\\
 $\dot{\psi_0}$ (mas/y) & Precession rate & -7608.3 & -- \\
 $\dot{I_0}$ (mas/y) & Obliquity rate & -2 & -- \\
 $\dot{\phi_0}$ (deg) & Rotation rate & 350.892 & fixed \\
 \hline
 $\phi_1^{cos}$ (mas) & Annual cosine spin angle & 481 & 70\\
 $\phi_2^{cos}$ (mas) & Semi-annual cosine spin angle & -103 & 25\\
 $\phi_3^{cos}$ (mas) & Ter-annual cosine spin angle & -35 & fixed\\
 $\phi_4^{cos}$ (mas) & 4-annual cosine spin angle & -10 & fixed\\
 $\phi_1^{sin}$ (mas) & Annual sine spin angle & -155 & 55 \\
 $\phi_2^{sin}$ (mas) & Semi-annual sine spin angle & -93 & 60\\
 $\phi_3^{sin}$ (mas) & Ter-annual sine spin angle & -3 & fixed\\
 $\phi_4^{sin}$ (mas) & 4-annual sine spin angle & -8 & fixed\\
 \hline
 $p_1$ (mas) & Annual prograde nutation & 102/104 & 20\\
 $p_2$ (mas) & Semi-annual prograde nutation & 498/512 & 19\\
 $p_3$ (mas) & Ter-annual prograde nutation & 108/112 & 12 \\
 $p_4$ (mas) & 4-annual prograde nutation & 18/18 & fixed\\
 $p_5$ (mas) & 5-annual prograde nutation & 3/3 & fixed\\
 $p_6$ (mas) & 6-annual prograde nutation & 0.5/0.5  & fixed\\
 $r_1$ (mas) & Annual retrograde nutation & 137/132 & 14\\
 $r_2$ (mas) & Semi-annual retrograde nutation & 18/15 & 21 \\
 $r_3$ (mas) & Ter-annual retrograde nutation & 5/12 & 21\\
 $r_4$ (mas) & 4-annual retrograde nutation & 1/1 & fixed \\
 $r_5$ (mas) & 5-annual retrograde nutation & 0/0 & fixed \\
 $r_6$ (mas) & 6-annual retrograde nutation & 0/0 & fixed \\
 \hline
 X (km) & X coordinate of the InSight / ExoMars lander & -2417 / 2879 & 200 \\
 Y (km) & Y coordinate of the InSight / ExoMars lander &  2366 / -1455 & 200 \\
 Z (km) & Z coordinate of the InSight / ExoMars lander &   266 / 1048 & 200 \\
  \hline

\end{tabular}
\caption{A priori Martian rotation model and a priori constraints used in the inversion procedure. The nutation amplitudes are given for the rigid case and for the fluid core (rigid/fluid). The model of rigid nutation is modulated by a liquid core of FCN period $T_{FCN} = -240$ days and core factor F = 0.07. $\phi^{cos,sin}_m$ are the amplitudes of the seasonal variations of the spin angle. 
$\dot{\psi_0}$ is the Mars precession rate of the spin axis, $\dot{I_0}$ is the secular change in the Mars obliquity and $\dot{\phi_0}$ is the Mars spin rate. $\psi_0$, $I_0$ and $\phi_0$ are the angle values at the J2000 epoch. }
\label{table:rotation_model}
\end{center}
\end{table}

\renewcommand{\arraystretch}{1.2}

\section{Simulations results}
\label{sec:res}
Here, we discuss only the estimated uncertainties for the semi-annual prograde $p_2$ and the ter-annual retrograde $r_3$ nutation amplitudes. $r_3$ is expected to be most amplified by the presence of the liquid core since it likely is close to the FCN period of Mars . We also include $p_2$ because it has the largest amplitude (\cite{dehant:2000}, \cite{lemaistre:2012}) and even a small alteration of $p_2$ already makes for detectable signature in the Doppler shift.
In the first subsection, we compare the sensitivities to the nutation amplitudes and the uncertainty level of their estimation for each mission alone. We assess the improvement of the nutation determination resulting from the combination of RISE and LaRa measurements in the second subsection. 

\subsection{Data sensitivity characterisation}
\label{ssec:sensitivity}

\begin{figure}[h!]
\begin{center} 
\includegraphics[scale=0.45]{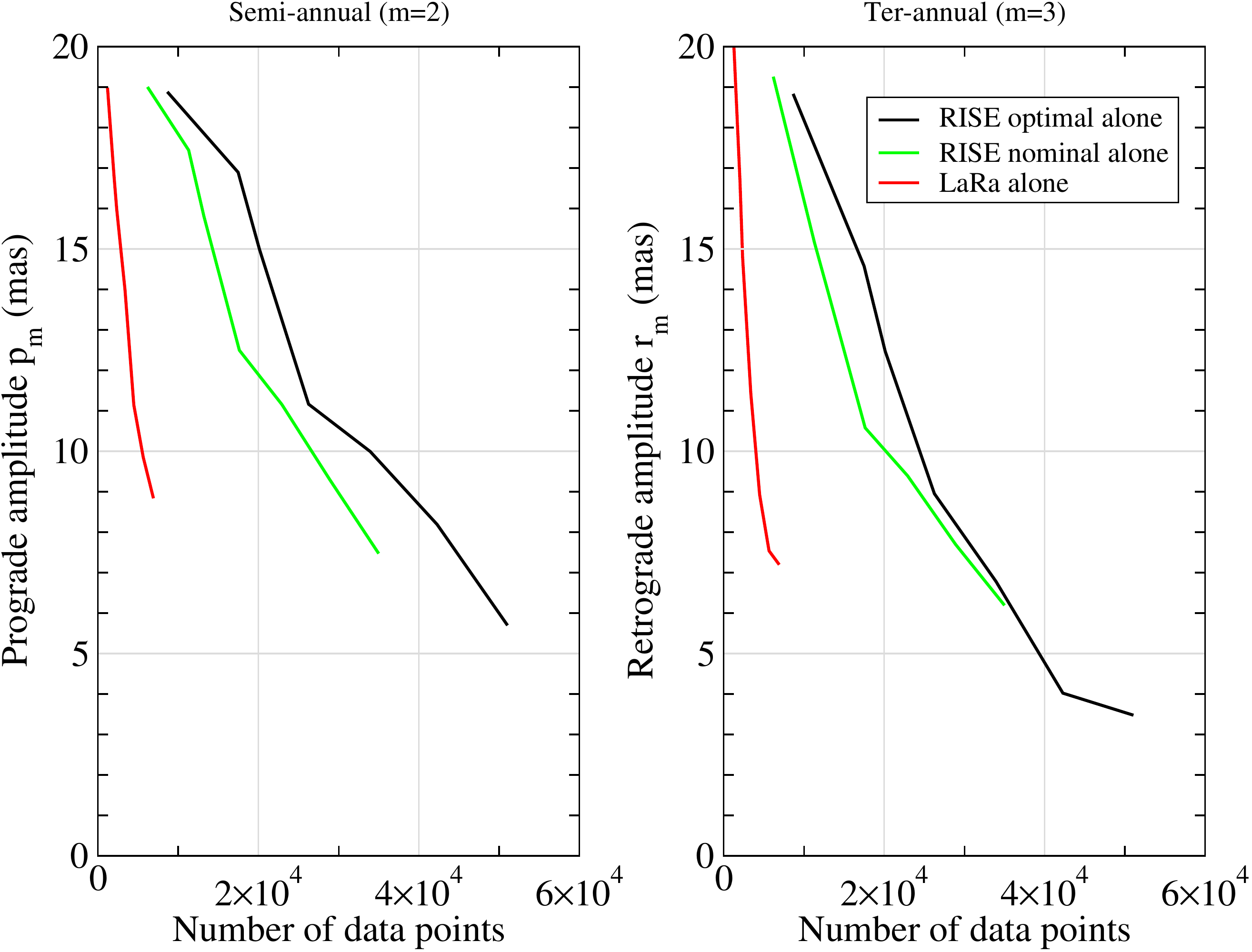}
\caption {Uncertainty (1$\sigma$) on $p_2$ and $r_3$ as a function of data points for the optimal and nominal RISE configurations and for the nominal LaRa configuration. The amplitudes are expressed in milliarcseconds (mas). 1~mas corresponds to 1.6 centimeters on Mars' surface. } \label{fig:data_points}
\end{center}
\end{figure}

\begin{figure}[h!]
\begin{center} 
\includegraphics[scale=0.45]{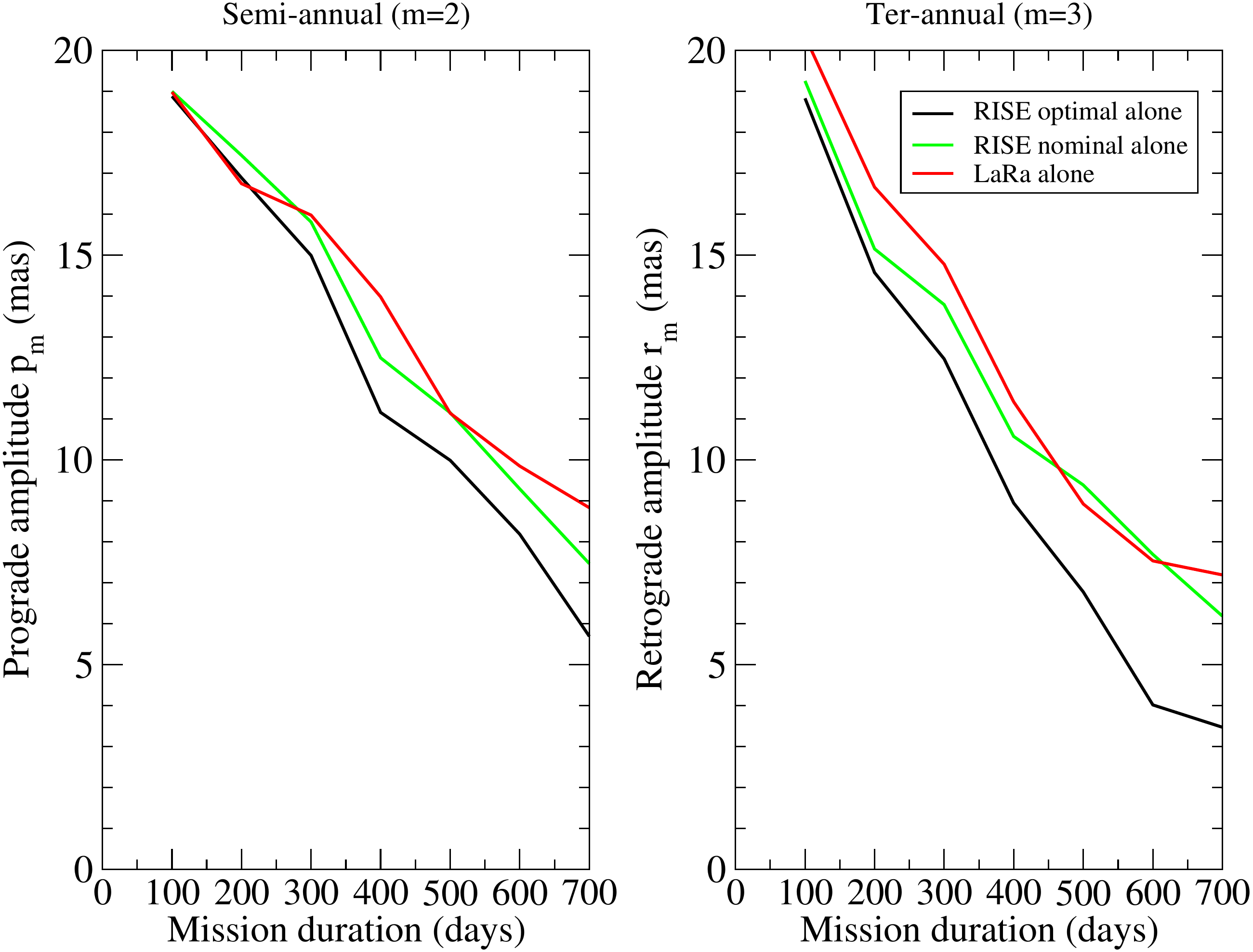}
\caption {Temporal evolution of the nutation uncertainties as estimated by RISE optimal case (in black), RISE nominal case (in green) and LaRa (in red) separately.} \label{fig:alone}
\end{center}
\end{figure}

Considering each mission separately, we estimate the expected precision on the semi-annual prograde $p_2$ and ter-annual retrograde $r_3$ nutation amplitudes as a function of the number of measurements points (see Fig.~\ref{fig:data_points}) and as a function of the mission duration (see Fig.~\ref{fig:alone}). We consider the two cases described in Section 2 for RISE: nominal (max. 25$\degree$ off bore sight and 5$\degree$ azimuthal offset) and optimal (alternate east-west tracking after 200 days and unconstrained azimuth).

The steeper slopes of the curves in Fig.~\ref{fig:data_points} clearly show the better efficiency of the LaRa experiment to determine the nutation parameters compared to RISE. Indeed, RISE needs about five to seven times more data to get the nutation amplitudes with a precision equivalent to LaRa's estimate. 
This is mostly due to the fact that a same number of data points covers a larger time period for LaRa than for RISE. Nevertheless, after 700 days and taking into account the dependence in the number of data N, we always have the uncertainty for LaRa $\sigma_{LaRa} < \sqrt{N_{RISE}/N_{LaRa}}~ \sigma_{RISE}$ for the nominal RISE case. This results from the richer geometry of LaRa in terms of azimuth allowed by the toroidal patterns of LaRa's monopole antennas. As discussed in Section 2, LaRa covers a wider range of azimuth and using data acquired from various LOS azimuths (and therefore at various local times of observation) has a strong positive impact on the estimated nutation uncertainties (see details in \cite{lemaistre:2019}). 
In the same way, the difference between the sensitivity of RISE nominal and RISE optimal solutions comes from the richer geometry of the latter scenario mostly due to the alternate tracking between eastward and westward directions.\\
\\
In Fig.~\ref{fig:alone}, we show that LaRa and RISE nominal separately provide quite similar precisions on the nutation estimate because the lower sensitivity of RISE is compensated by its higher number of measurement points. After 700 days of operation, $p_2$ and $r_3$ nutation amplitudes are estimated by LaRa alone with about 9 and 7 mas of uncertainty respectively, while RISE nominal alone would get $p_2$ with 7.5 mas uncertainty and $r_3$ with 6 mas uncertainty with 5 times more data points than LaRa. The RISE optimal scenario leads to uncertainties of about 6 and 3.5 mas for $p_2$ and $r_3$ respectively. 
\\
It is finally interesting to mention that we could have chosen to estimate the nutation transfer function parameters instead of the nutation amplitudes.
However, the nonlinear equations of the transfer function, whereas attractive because directly related to physical parameters and only dependent on two parameters F and $T_{FCN}$, cannot be solved properly by an iterative least squares process when the FCN period is close to the forcing period (Le Maistre et al. [2012]). This is the reason why we focus here on $p_2$ and $r_3$ that can be estimated with confidence with a least square method.
\\

\clearpage
\subsection{Synergy}
\label{ssec:synergy}

\begin{figure}[h!]
\begin{center} 
\includegraphics[scale=0.5]{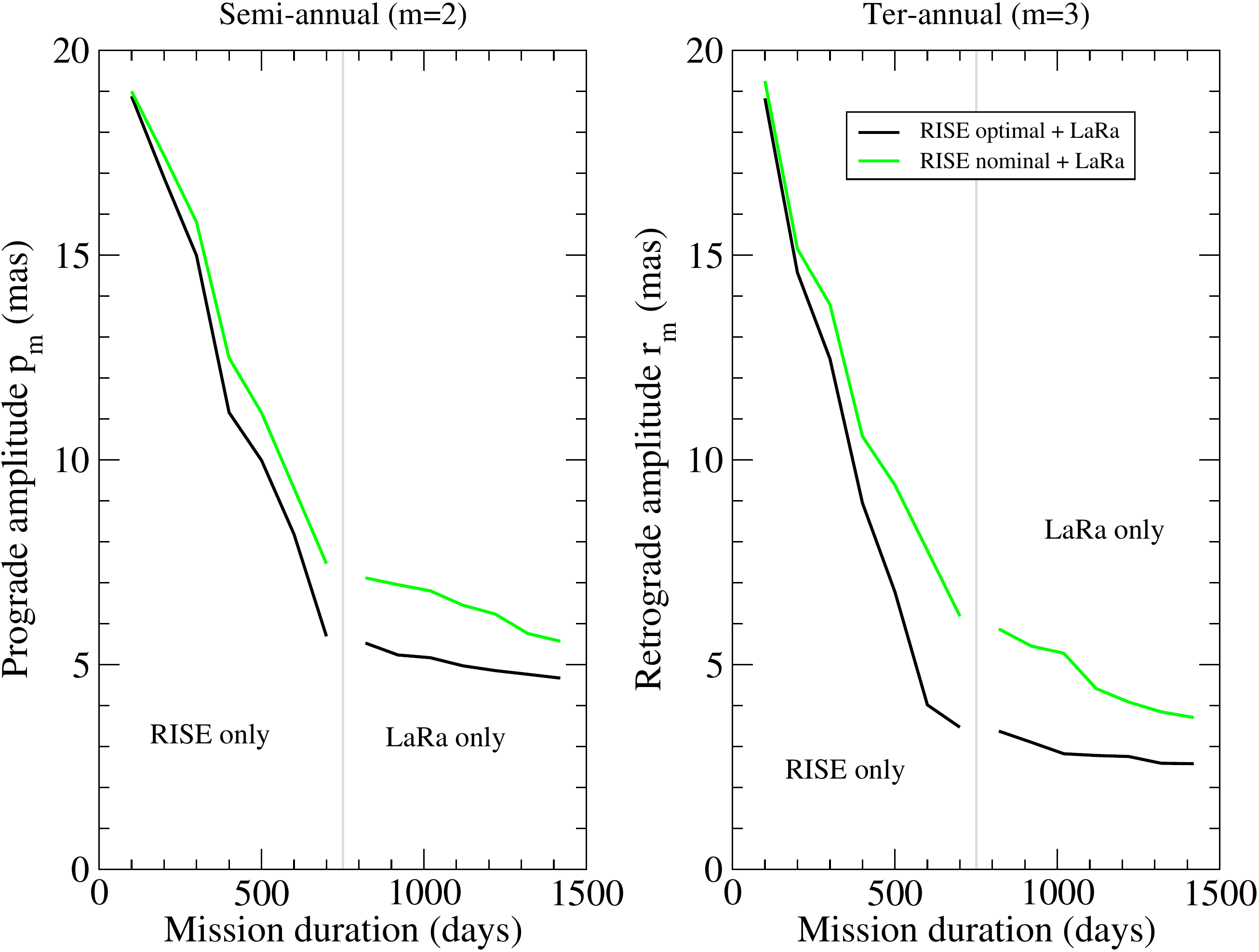}
\caption {Effect of combining RISE and LaRa on the $p_2$ and $r_3$ nutation uncertainties for the nominal RISE optimal case and the RISE nominal case followed by LaRa.} \label{fig:synergy}
\end{center}
\end{figure}

After discussing the expected precision on $p_2$ and $r_3$ for the two missions separately, we now show how the combination of the data from RISE and LaRa affects the precision on $p_2$ and $r_3$. Two scenarios are envisaged: RISE optimal case plus LaRa and RISE nominal case plus LaRa. 
To study the effect of combining both data sets, we stack the normal equations of the two missions to solve for the MOP (global parameters) and coordinates (local parameters) of both landers. We use the GINS software to produce the normal matrices and the DYNAMO software (also from CNES) to invert them and obtain the parameters estimate. \\
\\
The time evolution of the estimated uncertainties for the semi-annual prograde amplitude $p_2$ and the ter-annual retrograde amplitude $r_3$ are shown in Fig.~\ref{fig:synergy}. The estimated values of the MOP solutions and associated uncertainties from our simulations are given in Tab.~\ref{table:estimations} as well as true errors (difference between the estimated value and the value to retrieve). These results are consistent with \cite{lemaistre:2019} and \cite{dehant:2019} considering the slightly different set of estimated parameters and different settings described in \cite{dehant:2019}. After RISE tracking days (delineated by the grey vertical line), a precision of $\sim$ 7.5 mas and $\sim$ 6 mas for $p_2$ and $r_3$ respectively is obtained for the nominal case (in green). The combination with the LaRa tracking allows to further reduce the uncertainty on $p_2$ to 5.5~mas (i.e. 25\%), while the uncertainty on $r_3$ is reduced by about 35\% (i.e. up to 4~mas) after 1400 days. For the optimal RISE case (in black), the combination of RISE nominal with the LaRa tracking allows to determine $p_2$ with a precision of less than 5~mas and $r_3$ with about 2.5 mas uncertainty. The LaRa contribution will thus significantly help to recover from the adverse effect of off-pointing of the InSight antennas and the lack of variation in LOS azimuth direction of RISE. Especially in the more realistic nominal scenario, LaRa provides an important contribution for constraining the deep interior of Mars as discusses in the next section.\\
\\

\renewcommand{\arraystretch}{1.2}

\begin{table}[!t]
\begin{center}
\begin{tabular}{c c c c}
  \hline
  \hline
   \textbf{Symbol (unit)} & \textbf{Rotation parameter} & \textbf{MOP solution and uncertainty} &\textbf{True errors} \\
  \hline
   $\dot{\psi_0}$ (mas/y) & Precession rate & -7607.9 $\pm$ 0.4 & 0.4 \\
   $\dot{I_0}$ (mas/y) & Obliquity rate & -1.98 $\pm$ 0.2 & 0.1\\
  \hline
 $\phi_1^{cos}$ (mas) & Annual cosine spin angle &  468 $\pm$ 8 & 13\\
 $\phi_2^{cos}$ (mas) & Semi-annual cosine spin angle & -103 $\pm$ 8 & 0\\
 $\phi_1^{sin}$ (mas) & Annual sine spin angle & -147 $\pm$ 9 & 8\\
 $\phi_2^{sin}$ (mas) & Semi-annual sine spin angle & -99 $\pm$ 8 & 6\\  
\hline
 $p_1$ (mas) & Annual prograde nutation & 111 $\pm$ 6 & 7\\
 $p_2$ (mas) & Semi-annual prograde nutation & 511 $\pm$ 6 & 1\\
 $p_3$ (mas) & Ter-annual prograde nutation & 113 $\pm$ 2 & 1\\
 $r_1$ (mas) & Annual retrograde nutation & 134 $\pm$ 6 & 2\\
 $r_2$ (mas) & Semi-annual retrograde nutation & 19 $\pm$ 5 & 4\\
 $r_3$ (mas) & Ter-annual prograde nutation & 12 $\pm$ 4 & 0\\
  \hline

\end{tabular}
\caption{Estimated MOP mean values and uncertainties and true errors for the combination of RISE(nominal) and LaRa.}
\label{table:estimations}
\end{center}
\end{table}

\section{Implication for Mars interior}
\label{sec:interior}
Combining LaRa with RISE reduces the uncertainties in $p_2$ and $r_3$ by few mas with respect to RISE alone. We now discuss how those improved estimates affect inferences about the interior structure and in particular core size.\\
To address this point, we use a model of Mars' interior structure that agrees with its moments of inertia and global dissipation at the principal tidal period of Phobos to predict the nutations of the real Mars. Our visco-elastic hydrostatic model uses the EH45 mantle composition (\cite{sanloup:1999}) and a hot mantle end-member temperature profile deduced from thermal evolution studies (\cite{plesa:2016}). We assume a liquid convecting Fe-S core (\cite{rivoldini:2011}).\\

\begin{figure}[h!]
\noindent
\includegraphics[scale=0.35]{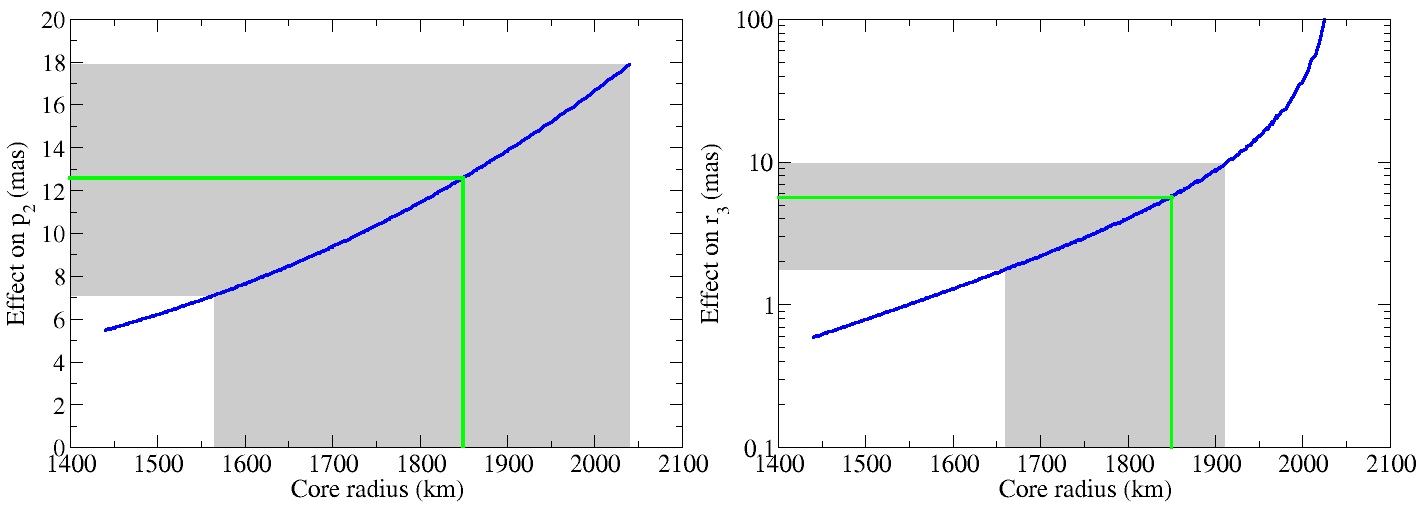}
 \caption {Amplification of the $p_2$ and $r_3$ because of the liquid core as a function of core radius and link between the nutation uncertainty and core radius uncertainty. The elastic hydrostatic models with the EH45 Mars mantle mineralogy and a hot mantle end-member temperature profile are considered. A liquid convecting Fe-S core is assumed. The grey region illustrates the nominal case for an average core size $r_{cmb}=1850$~km.}  \label{fig:tilio}
\end{figure}

Fig.~\ref{fig:tilio} shows the effect of the liquid core on the prograde semi-annual $p_2$ and the retrograde ter-annual $r_3$ nutation amplitudes as a function of core radius. The uncertainty on the core radius determination can be estimated from the expected uncertainties in $p_2$ and $r_3$ (Sec.~\ref{ssec:synergy}). For instance, for an assumed core radius of 1850 km \citep{plesa:2018} and the expected uncertainties for RISE nominal combined with LaRa of about 5.5 mas on $p_2$ and 4 mas on $r_3$, the core radius can be determined with a precision of about 260 and 125 km (see Fig. 8).  In Table ~\ref{table:interp}, we show the resulting uncertainties in $r_{cmb}$ for the two scenarios (nominal and optimal synergies) and for four different core radii $r_{cmb}$ (1450, 1600, 1850, 2000 km). 
 
$p_2$ will allow confirming the liquid state of the core, but constraints on its size will be deduced mostly from $r_3$ because it is expected that the period of the FCN is closest to the period of the ter-annual nutation and therefore has the largest effect of the liquid core on $r_3$. As a consequence, the resulting precision on the core radius, if determined from $r_3$, depends on the FCN period of Mars, i.e. the closer the period of the FCN is to the period of $r_3$, the more precisely the core radius can be estimated.

A more detailed analysis of core properties that can be determined from nutation observations requires considering additional Martian interior structures than considered here, such as different possible mantle mineralogies, and will increase the inferred uncertainties more than 10\% (e.g. \cite{rivoldini:2019aa}). Also Mars' non-hydrostatic state modifies the nutations through its effect on the shape of the core-mantle boundary, and can lenghten the FCN, possibly bringing the FCN period closer to the $r_3$ nutation period (\cite{wieczorek:2019}, \cite{rivoldini:2019aa}), which would be beneficial for the precision on the core radius from $r_3$, as illustrated by Table 4.

The level of uncertainty in $r_{cmb}$ is very sensitive to the nutation amplitude level of uncertainty. Therefore, LaRa alone will not improve significantly the results from RISE (see Fig.~\ref{fig:synergy}) but we can expect a more precise estimate of $r_{cmb}$ with the data from both landers together. Additionally LaRa will provide an independent confirmation of the RISE results.
Note if the core of Mars is not in hydrostatic equilibrium, the FCN could be shifted (\cite{rivoldini:2019aa}) and the location of the resonance on Fig. 8 could therefore also change.

\renewcommand{\arraystretch}{1.1}

\begin{table}[t!]
\begin{center}
\begin{tabular}{c c c c c}
  \hline
   & \multicolumn{4}{c}{\textbf{Uncertainties on the core radius}}  \\
  \hline
   \textbf{$r_{cmb}$} (km) & \multicolumn{2}{c}{\textbf{Synergy optimal case (km)}} & \multicolumn{2}{c}{\textbf{Synergy nominal case (km)}}  \\
   & \multicolumn{1}{c}{$p_2$} & \multicolumn{1}{c}{$r_3$}  & \multicolumn{1}{c}{$p_2$} & \multicolumn{1}{c}{$r_3$}\\
   & \multicolumn{1}{c}{(\footnotesize{5 mas})} & \multicolumn{1}{c}{(\footnotesize{2.5 mas})} & \multicolumn{1}{c}{(\footnotesize{5.5 mas})} & \multicolumn{1}{c}{(\footnotesize{4 mas})} \\
  \hline
   1440 & + 315 & + 320 & + 360 & + 380 \\
   1600 & $\pm$ 205 & $\pm$ 175 & $\pm$ 225 & $\pm$ 200 \\
   1850 & $\pm$ 218 & $\pm$ 65 & $\pm$ 260 & $\pm$ 125 \\
   2000\footnotemark[2] & -195/+45 & $\pm$ 5 & -235/+45 & $\pm$ 10\\
 \hline

\end{tabular}
\caption{Inferred uncertainties on the core radius for different values of the core radius itself, as deduced from the nutation uncertainties of Tab.~\ref{table:estimations} in nominal and optimal synergy cases.}
\label{table:interp}
\end{center}
\end{table}

\section{Conclusion}
\label{sec:conclusion}
This paper demonstrates the benefit of having LaRa after RISE for the determination of the Mars orientation and rotation parameters. Our simulations show that LaRa will reduce the MOP estimates uncertainties with respect to the uncertainty level expected from RISE alone. By combining 700 nominal RISE tracking days with 700 LaRa tracking days, we will achieve a precision of 5.5 mas and 4~mas for the semi-annual prograde and the ter-annual retrograde nutations respectively. Although LaRa will operate 5 to 7 times less than RISE, LaRa has the advantage of its richer geometry which compensates the limited amount of data. The combination allows then to reduce the uncertainties on $p_2$ and $r_3$ by about 25$\%$ and 35$\%$ respectively.

Finally, according to our simulations, combining the data provided by both instruments clearly improve the expected knowledge about the core. Even if RISE is a total success, the contribution of LaRa (even if small) in MOP determination can have a significant impact on the constraints inferred on the interior structure (e.g. core size). Additionally LaRa will provide an independent confirmation of the RISE results. The above expected uncertainties on $p_2$ and $r_3$ could lead to uncertainties of about 260 and 125 km on the core radius, assuming a core radius $r_{cmb}$ = 1850 km and could pinpoint $r_{cmb}$ with 10 km uncertainty if the FCN period is close to a forcing period (resonance).
\footnotetext[2]{The negative part is constrained by the nutations and the positive part is constrained by the precession linked to MoI.}

\section*{Acknowledgments}
This work was financially supported by the Belgian PRODEX program managed by the European Space Agency in collaboration with the Belgian Federal Science Policy Office. This is InSight contribution number ICN 122.
\clearpage
\bibliography{ref}

\begin{thebibliography}{25}
\providecommand{\natexlab}[1]{#1}
\providecommand{\url}[1]{\texttt{#1}}
\expandafter\ifx\csname urlstyle\endcsname\relax
  \providecommand{\doi}[1]{doi: #1}\else
  \providecommand{\doi}{doi: \begingroup \urlstyle{rm}\Url}\fi

\bibitem[Defraigne et~al.(1995)Defraigne, Dehant, and
  P{\^a}quet]{defraigne:1995}
P.~Defraigne, V.~Dehant, and P.~P{\^a}quet.
\newblock Link between the retrograde-prograde nutations and nutations in
  obliquity and longitude.
\newblock \emph{Celestial Mechanics and Dynamical Astronomy}, 62\penalty0
  (4):\penalty0 363--376, 1995.

\bibitem[Dehant and Mathews(2015)]{dehant:2015aa}
V.~Dehant and P.M. Mathews.
\newblock \emph{Precession, Nutation and Wobble of the Earth}.
\newblock Cambridge University Press, Cambridge, 2015.

\bibitem[Dehant et~al.(2000b)Dehant, Van~Hoolst, and Defraigne]{dehant:2000}
V.~Dehant, T.~Van~Hoolst, and P.~Defraigne.
\newblock {Comparison Between the Nutations of the Planet Mars and the
  Nutations of the Earth}.
\newblock \emph{Surveys in Geophysics}, 21:\penalty0 89--110, 2000b.

\bibitem[Dehant et~al.(2019)Dehant, Le~Maistre, Baland, Karatekin, P\'eters,
  Rivoldini, Van~Hoolst, Yseboodt, Mitrovic, and the LaRa~team]{dehant:2019}
V.~Dehant, S.~Le~Maistre, R.M. Baland, O.~Karatekin, M.-J. P\'eters,
  A.~Rivoldini, T.~Van~Hoolst, M.~Yseboodt, M.~Mitrovic, and the LaRa~team.
\newblock {The radio-science LaRa instrument onboard ExoMars 2020 to
  investigate the rotation and interior of Mars}.
\newblock \emph{Manuscript in preparation}, 2019.

\bibitem[Folkner et~al.(1997)Folkner, Yoder, Yuan, Standish, and
  Preston]{folkner:1997}
W.M. Folkner, C.F. Yoder, D.N. Yuan, E.M. Standish, and R.A. Preston.
\newblock {Interior structure and seasonal mass redistribution of Mars from
  radio tracking of Mars Pathfinder}.
\newblock \emph{Science}, 278\penalty0 (5344):\penalty0 1749--1752, 1997.

\bibitem[Folkner et~al.(2018)Folkner, Dehant, Le~Maistre, Yseboodt, Rivoldini,
  Van~Hoolst, Asmar, and Golombek]{Folkner:2018}
W.M. Folkner, V.~Dehant, S.~Le~Maistre, M.~Yseboodt, A.~Rivoldini,
  T.~Van~Hoolst, S.W. Asmar, and M.P Golombek.
\newblock {The rotation and interior structure experiment on the InSight
  mission to Mars}.
\newblock \emph{Space Science Reviews}, 214\penalty0 (5):\penalty0 100, 2018.

\bibitem[Karki et~al.(2019)Karki, Sabbadini, Alkhalifeh, and
  Craeye]{Karki:2019aa}
S.~Karki, M.~Sabbadini, K.~Alkhalifeh, and C.~Craeye.
\newblock {Metallic Monopole Parasitic Antenna with Circularly Polarized
  Conical Patterns}.
\newblock \emph{IEEE Transactions on Antennas and Propagation}, 2019.
\newblock \doi{10.1109/TAP.2019.2916737}.

\bibitem[Konopliv et~al.(2016)Konopliv, Park, and Folkner]{konopliv:2016}
A.S. Konopliv, R.S. Park, and W.M. Folkner.
\newblock {An improved JPL Mars gravity field and orientation from Mars orbiter
  and lander tracking data}.
\newblock \emph{Icarus}, 274:\penalty0 253--260, 2016.
\newblock \doi{https://doi.org/10.1016/j.icarus.2016.02.052}.

\bibitem[Kuchynka et~al.(2014)Kuchynka, Folkner, Konopliv, Parker, Park,
  Le~Maistre, and Dehant]{kuchynka:2014}
P.~Kuchynka, W.M. Folkner, A.S. Konopliv, T.J. Parker, R.S. Park,
  S.~Le~Maistre, and V.~Dehant.
\newblock {New constraints on Mars rotation determined from radiometric
  tracking of the Opportunity Mars Exploration Rover}.
\newblock \emph{Icarus}, 229:\penalty0 340--347, 2014.

\bibitem[Le~Maistre(2013)]{lemaistre:2013}
S.~Le~Maistre.
\newblock \emph{{The rotation of Mars and Phobos from Earth-based
  radio-tracking observations of a lander}}.
\newblock PhD thesis, Universit{\'e} Catholique de Louvain, November 2013.

\bibitem[Le~Maistre et~al.(2012)Le~Maistre, Rosenblatt, Rivoldini, Dehant,
  Marty, and Karatekin]{lemaistre:2012}
S.~Le~Maistre, P.~Rosenblatt, A.~Rivoldini, V.~Dehant, J.-C. Marty, and
  O.~Karatekin.
\newblock {Lander radio science experiment with a direct link between Mars and
  the Earth}.
\newblock \emph{Planetary and Space Science}, 68\penalty0 (1):\penalty0
  105--122, 2012.

\bibitem[Le~Maistre et~al.(2019)Le~Maistre, P\'eters, J.-C., Dehant, and the
  LaRa~team]{lemaistre:2019}
S.~Le~Maistre, M.-J. P\'eters, Marty J.-C., V.~Dehant, and the LaRa~team.
\newblock {On the impact of the operational and technical characteristics of
  the LaRa experiment on the nutation determination}.
\newblock \emph{Manuscript in preparation}, 2019.

\bibitem[{Parker} et~al.(2019){Parker}, {Golombek}, {Calef}, {Williams}, {Le
  Maistre}, {Folkner}, {Daubar}, {Kipp}, {Sklyanskiy}, {Lethcoe-Wilson}, and
  {Hausmann}]{Parker:2019aa}
T.~J. {Parker}, M.~P. {Golombek}, F.~J. {Calef}, N.~R. {Williams}, S.~{Le
  Maistre}, W.~{Folkner}, I.~J. {Daubar}, D.~{Kipp}, E.~{Sklyanskiy},
  H.~{Lethcoe-Wilson}, and R.~{Hausmann}.
\newblock {Localization of the InSight Lander}.
\newblock In \emph{Lunar and Planetary Science Conference}, Lunar and Planetary
  Science Conference, page 1948, Mar 2019.

\bibitem[Plesa et~al.(2016)Plesa, Grott, Tosi, Breuer, Spohn, and
  Wieczorek]{plesa:2016}
A.-C. Plesa, M.~Grott, N.~Tosi, D.~Breuer, T.~Spohn, and M.A. Wieczorek.
\newblock {How large are present-day heat flux variations across the surface of
  Mars?}
\newblock \emph{Journal of Geophysical Research: Planets}, 121\penalty0
  (12):\penalty0 2386--2403, 2016.

\bibitem[{Plesa} et~al.(2018){Plesa}, {Padovan}, {Tosi}, {Breuer}, {Grott},
  {Wieczorek}, {Spohn}, {Smrekar}, and {Banerdt}]{plesa:2018}
A.~C. {Plesa}, S.~{Padovan}, N.~{Tosi}, D.~{Breuer}, M.~{Grott}, M.~A.
  {Wieczorek}, T.~{Spohn}, S.~E. {Smrekar}, and W.~B. {Banerdt}.
\newblock {The Thermal State and Interior Structure of Mars}.
\newblock \emph{Geophysical Research Letters}, 45\penalty0 (22):\penalty0
  12,198--12,209, Nov 2018.
\newblock \doi{10.1029/2018GL080728}.

\bibitem[Rivoldini et~al.(2011)Rivoldini, Van~Hoolst, Verhoeven, Mocquet, and
  Dehant]{rivoldini:2011}
A.~Rivoldini, T.~Van~Hoolst, O.~Verhoeven, A.~Mocquet, and V.~Dehant.
\newblock {Geodesy constraints on the interior structure and composition of
  Mars}.
\newblock \emph{Icarus}, 213\penalty0 (2):\penalty0 451--472, 2011.

\bibitem[Rivoldini et~al.(2019)Rivoldini, Beuthe, Van~Hoolst, Wieczorek,
  Baland, Dehant, Folkner, Gudkova, Le~Maistre, Peters, Yseboodt, and
  Zharkov]{rivoldini:2019aa}
A.~Rivoldini, M.~Beuthe, T.~Van~Hoolst, M.~Wieczorek, R-M. Baland, V.~Dehant,
  B.~Folkner, T.~Gudkova, S.~Le~Maistre, M-J. Peters, M.~Yseboodt, and
  V.~Zharkov.
\newblock {Non-hydrostatic effects on Mars’ nutation}.
\newblock In \emph{EGU General Assembly Conference Abstracts}, volume~21, page
  17999, Apr 2019.

\bibitem[Roosbeek(1999)]{Roosbeek:1999}
F.~Roosbeek.
\newblock {Analytical developments of rigid Mars nutation and tide generating
  potential series}.
\newblock \emph{Celestial Mechanics and Dynamical Astronomy}, 75\penalty0
  (4):\penalty0 287--300, 1999.

\bibitem[Sanloup et~al.(1999)Sanloup, Jambon, and Gillet]{sanloup:1999}
C.~Sanloup, A.~Jambon, and P.~Gillet.
\newblock A simple chondritic model of mars.
\newblock \emph{{Physics of the Earth and Planetary Interiors}}, 112\penalty0
  (1-2):\penalty0 43--54, 1999.

\bibitem[Sasao et~al.(1980)Sasao, Okubo, and Saito]{sasao:1980}
T.~Sasao, S.~Okubo, and M.~Saito.
\newblock {A Simple Theory on the Dynamical Effects of a Stratified Fluid Core
  Upon Nutational Motion of the Earth}.
\newblock \emph{Symposium - International Astronomical Union}, 78:\penalty0
  165–183, 1980.

\bibitem[Van Den~Acker et~al.(2002)Van Den~Acker, Van~Hoolst, de~Viron,
  Defraigne, Forget, Hourdin, and Dehant]{vandenacker:2002}
E.~Van Den~Acker, T.~Van~Hoolst, O.~de~Viron, P.~Defraigne, F.~Forget,
  F.~Hourdin, and V.~Dehant.
\newblock {Influence of the seasonal winds and the CO2 mass exchange between
  atmosphere and polar caps on Mars' rotation}.
\newblock \emph{Journal of Geophysical Research: Planets}, 107\penalty0 (E7),
  2002.

\bibitem[Van~Hoolst(2015)]{vanhoolst:2015}
T.~Van~Hoolst.
\newblock Rotation of the terrestrial planets.
\newblock pages 121--151, 01 2015.

\bibitem[{Wieczorek} et~al.(2019){Wieczorek}, {Beuthe}, {Rivoldini}, and {Van
  Hoolst}]{wieczorek:2019}
Mark~A. {Wieczorek}, Mikael {Beuthe}, Attilio {Rivoldini}, and Tim {Van
  Hoolst}.
\newblock {Hydrostatic Interfaces in Bodies With Nonhydrostatic Lithospheres}.
\newblock \emph{Journal of Geophysical Research (Planets)}, 124\penalty0
  (5):\penalty0 1410--1432, May 2019.
\newblock \doi{10.1029/2018JE005909}.

\bibitem[Yoder and Standish(1997)]{yoder:1997}
C.F. Yoder and E.M. Standish.
\newblock {Martian precession and rotation from Viking lander range data}.
\newblock \emph{Journal of Geophysical Research: Planets}, 102\penalty0
  (E2):\penalty0 4065--4080, 1997.

\bibitem[Yseboodt et~al.(2017)Yseboodt, Dehant, and Péters]{yseboodt:2017}
M.~Yseboodt, V.~Dehant, and M.-J. Péters.
\newblock {Signatures of the Martian rotation parameters in the Doppler and
  range observables}.
\newblock \emph{Planetary and Space Science}, 144:\penalty0 74--88, 2017.

\end{thebibliography}
\bibliographystyle{plainnat}

\end{document}